\documentclass[preprint,prX,amsmath,amssymb]{revtex4}
\usepackage{graphicx}
\usepackage{amssymb,amsfonts,amsmath}

\begin{document}



\title{Context-dependent interaction leads to emergent search behavior in social aggregates}




\author{Colin Torney, Zoltan Neufeld, and Iain D. Couzin\\}




\begin{abstract}
\par
Locating the source of an advected chemical signal is a common challenge facing many living organisms. When the advecting medium is characterized by either high Reynolds number or high Peclet number the task becomes highly non-trivial due to the generation of heterogenous, dynamically changing filamental concentrations which do not decrease monotonically with distance to the source. Defining search strategies which are effective in these environments has important implications for the understanding of animal behavior and for the design of biologically inspired technology. Here we present a strategy which is able to solve this task without the higher intelligence required to assess spatial gradient direction, measure the diffusive properties of the flow field or perform complex calculations. Instead our method is based on the collective behavior of autonomous individuals following simple social interaction rules which are modified according to the local conditions they are experiencing. Through these context-dependent interactions the group is able to locate the source of a chemical signal and in doing so displays an awareness of the environment not present at the individual level. Our model demonstrates the ability of decentralized information processing systems to solve real world problems and also illustrates an alternative pathway to cooperative behavior and evolution of higher cognitive capacity via the emergent, group level intelligence which can result from local interactions.
\end{abstract}


\maketitle




Throughout the natural world organisms are constantly faced with the challenge of locating the resources required for their survival. Often this means navigating their environment based on spatiotemporally varying information such as advected chemical cues, thermal gradients or magnetic fields. It has been noted collective behavior can greatly assist animal navigation. One explanation for this, known as the `many wrongs' principle \cite{wrongs}, is that inherent noise in the environment is dampened due to multiple sampling by individuals within a group. A quantitative study of an effect of this type was made by Gr\"{u}nbaum \cite{grun} and the benefits of sociality clearly illustrated. However this effect does not capture the potential emergent properties of social aggregations which often display complex behaviors not possible at the individual level \cite{complex}, are able to effectively store and process information, and make accurate consensus decisions in the absence of explicit communication \cite{cama, couz}. In this context complex systems, such as fish schools or animal herds, can be viewed as information processing entities with a collective awareness of their environment. Understanding their capacity for performing search tasks will not only shed light on the evolutionary pressures leading to aggregation but may also have important consequences for the development of distributed technologies such as olfactory robot swarms with applications in the detection of explosives, landmines, or locating people in search and rescue operations \cite{robot, robots2}. 

\par
The use of advected chemical signals by organisms in order to gain information about their environment is a ubiquitous behavior commonly seen in aquatic animals or flying insects and observed in a diverse range of species from the bee hunting wasp, \textit{Philanthus} \cite{tin} to parasitic plants able to grow towards a potential host \cite{plant}. The exact mechanisms which allow organisms to respond effectively to these signals is poorly understood. This is particularly true when signals are advected by stochastically fluctuating, chaotic flows. This issue has been addressed previously and algorithms based on a statistical \cite{shraiman} or information theoretic \cite{vergassola} approach have been developed. Here we consider an approach requiring less cognitive or sensing capacity on the part of an individual but instead relies on interactions with conspecifics to generate a search response which is effective in tracking an advected chemical filament. The key mechanism, which may be generalized to other situations, lies in the continual adjustment of an individual's behavior, from being more or less independent to entirely following its neighbors, as a function on its level of confidence in its own current strategy. Consequently the leadership structure of the group changes, continuously adapting to the quality of local information available to the members. For the search problem the individual strategy is the presumed direction towards the target along the concentration filament and confidence is assessed based on the changes in the concentration of the olfactory signal sensed along the trajectory in the recent past.
\par
It is assumed that the transport properties of the flow considered exhibit characteristics observed in chaotic advection which lead to thin filaments of chemical concentration in which the steepest gradients are transverse to the direction of the source and density is non-monotonically decreasing with distance to the source location (i.e. patches of high concentration are advected downstream). The dynamics of the chemical field is represented by
\begin{equation}
\frac{\partial C }{\partial t}  +  \mathbf{v_{f}} \cdot \nabla C
 =  S(\mathbf{r}_0) - bC
\end{equation}
where $S(\mathbf{r}_0)$ is a constant source at a point in space, $b$ is a decay rate and $\mathbf{v_{f}}$ is a stochastic velocity field with an imposed mean flow along the $y$-axis. The stochastic field is generated by the random evolution in Fourier space of a prescribed energy spectrum \cite{torney, careta} with an exponential decay. 
\begin{figure*}[t]
\includegraphics[scale=0.85,clip]{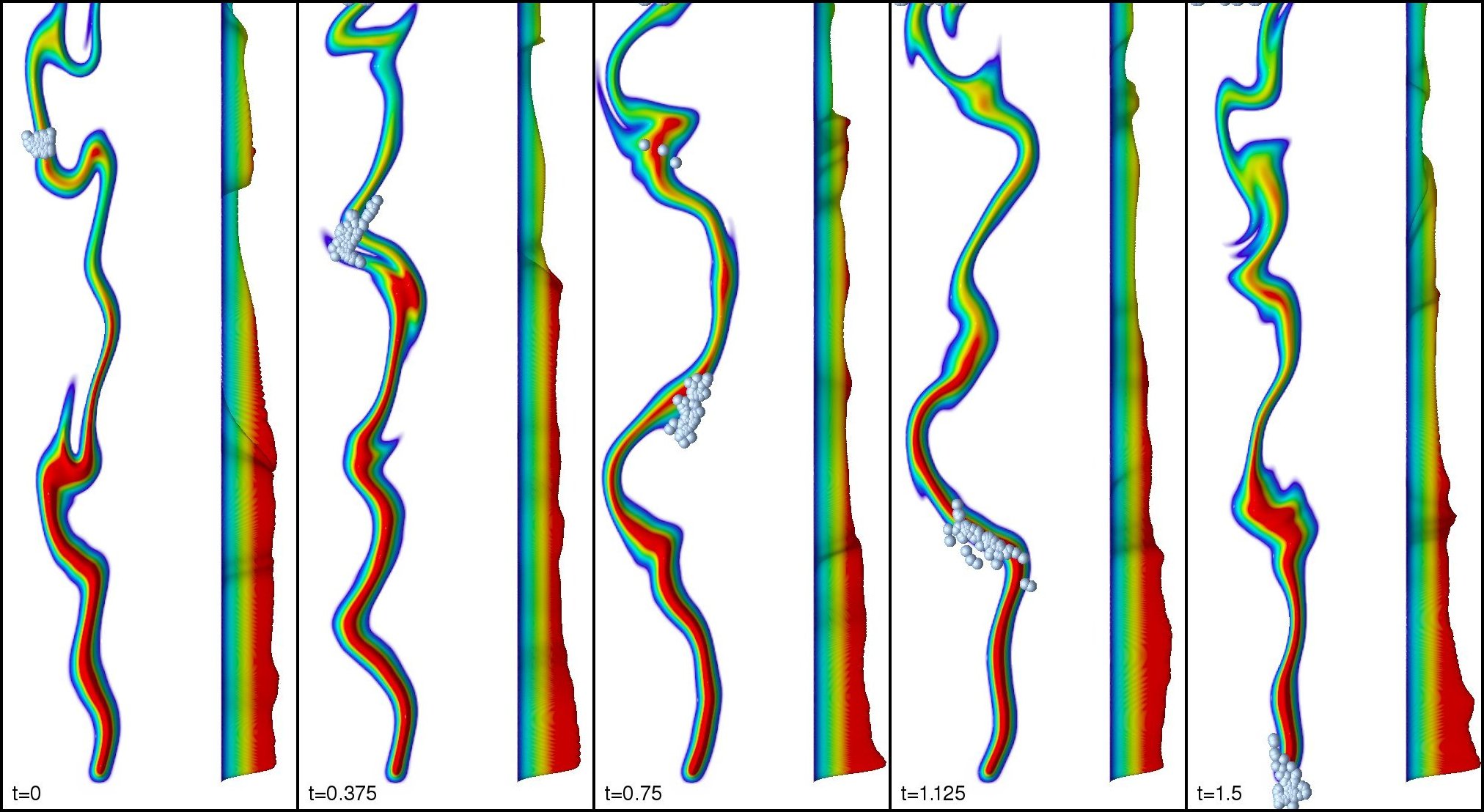}
\caption{Snapshots of 60 individuals performing filament tracking at intervals of 0.375, 53 successfully locate the source, repulsion zone size $2\times10^{-3}$, $\alpha=12.5\times10^{-3}$. Top view of filament and individuals is shown on left of each sequence, while on the right is the rotated concentration profile.}
\label{fig1}
\end{figure*}

\section{Interaction rules}
Individuals in our model are advected by the flow and move at a constant speed in a direction defined by their orientation
\begin{equation}
\mathbf{\dot{r}}_i = \mathbf{v_{f}}(\mathbf{r}_i,t) + v_s \mathbf{p}_i
\end{equation}
The flow is simulated in a domain of length L and a timescale for the model is selected by relating this length to a characteristic velocity of the model, the absolute velocity of an individual if it is perfectly aligned against the imposed mean flow. Therefore $T=L/(v_s-{\bar{v}_{fy}})$ i.e. an individual moving against the mean flow will cross the domain in unit time. All further parameters are then defined in terms of these units.
\par
The time evolution of the orientation vector results from interaction rules based on an abstraction of aggregation tendencies observed in biological systems. A range of theoretical models have been introduced to describe the collective motion of animal groups (see e.g. \cite{vicsek, gueron, mogilner}); here we follow the approach of Aoki \cite{aoki} and Huth \& Wissel \cite{huth} in assuming that preferred direction results from local alignment, repulsion and attraction, however we assume the interaction zone over which an individual responds to neighbours varies as a function of current local conditions. This follows a similar concept outlined in \cite{huth2} which demonstrated simulated fish schools were able to track regions of improved abiotic conditions based on modifying reactions to conspecifics and reducing speed when located in preferred regions.
\par
The direction of motion $\mathbf{p}_i$ is modified on the basis of the position and velocity of neighbouring individuals. A desired direction $\mathbf{d}_i$ is defined by three social interaction rules. In order to maintain a minimum distance between neighbours individuals move away from those within a repulsion zone $Z_R$ (this precludes any other behavioral response) 
\begin{equation}
\mathbf{d}_i= -\sum\limits_{j \in Z_R} \frac{\mathbf{r}_j-\mathbf{r}_i}{\vert{\mathbf{r}_j-\mathbf{r}_i}\vert }.
\end{equation}
If no others are present within the repulsion zone individuals move towards those within an attraction zone $Z_A$ and seek to align themselves with their neighbours within an orientation zone $Z_O$
\begin{equation}
\mathbf{d}_i= \sum\limits_{j \in Z_A} \frac{\mathbf{r}_j-\mathbf{r}_i}{\vert{\mathbf{r}_j-\mathbf{r}_i}\vert } + \sum\limits_{j \in Z_O}\mathbf{p}_j
\end{equation}
where the attraction zone is typically assumed to be larger than the alignment zone. Each individual turns towards $\mathbf{d}_i$ at a rate proportional to the difference between the current and desired direction with a maximum angular velocity defined by a parameter $\gamma$. If no neighbours are present within the interaction zone the direction vector $\mathbf{p}_i$ remains unchanged.

\section{Context-dependence}
In order to couple environment to behavior and therefore create an effective search algorithm, each individual varies the size of the interaction zones of local alignment and attraction depending on the current concentration value experienced. It should be noted that this is not a directional measure, no gradient is detected and flow velocity is ignored, therefore at the level of a single
individual a search strategy does not even exist.
\par
This model is not aimed to be an accurate representation of the behavior of a particular species, and a range of other qualitatively similar responses could be employed by different types of animal groups and in various environments. For example, varying the turning rates or some weighting factors of the alignment, attraction and actual direction as a function of some measure of confidence can lead to qualitatively similar results.  It has been shown experimentally that foraging success for groups of schooling fish does not linearly improve with group size \cite{steele} as would be expected by the `many wrongs' principle and it can therefore be assumed that this is a result of some non-trivial interactions between individuals, however these experiments are inconclusive and further investigation is required. While the exact nature of interactions can never be fully determined, simulations where the size of interaction zones are varied based on local conditions are able to accurately model observed experimental results \cite{context}. Apart from its biological relevance the model presented here also provides an effective and robust algorithm with potential applications for distributed, autonomous systems.
\par
The kernel of the search behavior therefore lies in the expansion and contraction of the attraction and orientation zones based on the local concentration experienced by an individual. The radius of each zone is a function of a normalized individual concentration parameter $C_i\left(t\right)\in [0,1]$ that measures the confidence in the actual direction of motion based on the recently encountered signal, defined as
\begin{equation}
{C_i}\left(t\right)=\frac  {C\left(\mathbf{r}_i, t\right)}{\max\limits_{0< \tau < t} \left[ C\left(\mathbf{r}_i\left(t-\tau\right), t-\tau \right) e^{-\tau/\alpha}\right]}.
\end{equation}
The denominator is a record of the history of the individual's trajectory, while $\alpha$ controls the decay time and acts as a short-term memory which allows individuals to assess the local concentration in the context of recent experience only. 
Note, that this type of temporal comparison of sensed concentration values is also an essential component of the
well known run-and-tumble model of bacterial chemotaxis \cite{segall}.
\par

\begin{figure}[t]
\includegraphics[angle=-90,scale=0.33,clip]{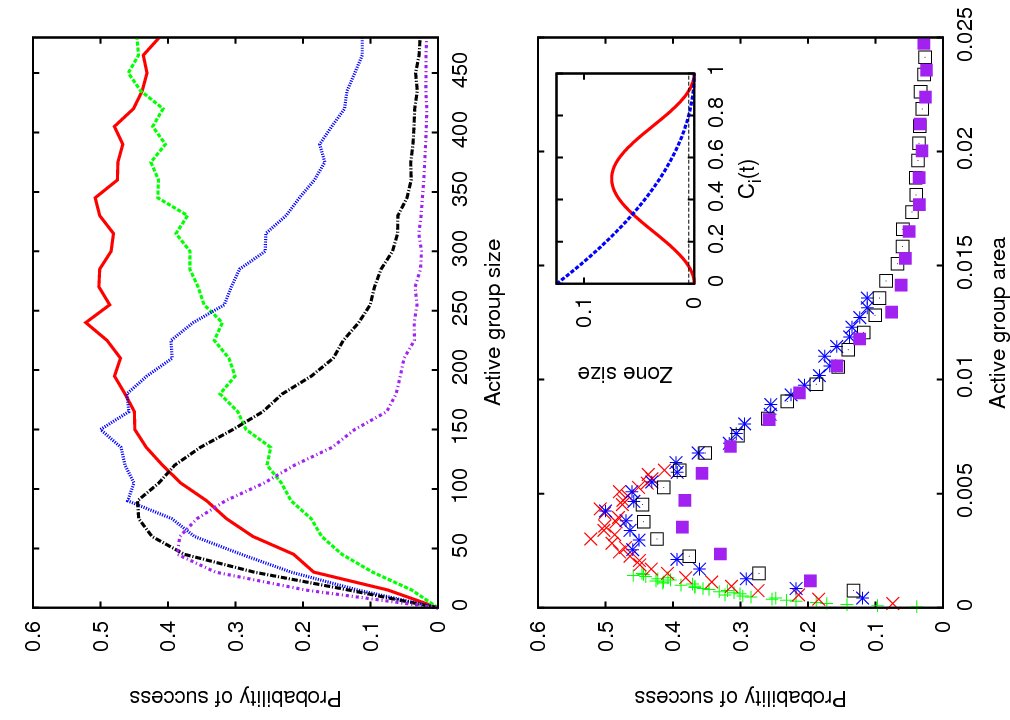}
\caption{(a) Active group size vs. probability of successfully locating the source. Lines represent varying repulsion zone size, green dash $1\times10^{-3}$, red solid $2\times10^{-3}$, blue dot $3\times10^{-3}$, black dot-dash $4\times10^{-3}$, purple dot-small dash $5\times10^{-3}$. (b) Active group area vs success rate green + $1\times10^{-3}$, red x $2\times10^{-3}$, blue $*$ $3\times10^{-3}$, black square (outline) $4\times10^{-3}$, purple square (solid) $5\times10^{-3}$; inset interaction zone sizes as a function of $C_i(t)$, red dash attraction zone, blue solid orientation zone, black dash repulsion zone}
\label{fig2}
\end{figure}

In our two-dimensional model an individual controls the radius of the attraction and orientation zone it responds to by applying a scaling factor to the maximum size of these values ($R_A$ and $R_O$ respectively) dependent on the concentration parameter. The radius of interaction for each response are then defined as (see inset Fig.~\ref{fig2}b)
\begin{eqnarray}
R_A(t) = (1-C_i\left(t\right))^2 R_{A,max}\nonumber \\
R_O(t) = \sin^2(\pi C_i\left(t\right)) R_{O,max}
\end{eqnarray}
Although the exact functional form of this relation is reasonably arbitrary the key behavioral response they represent can be summarized as \begin{itemize}
\item a sharp decrease in signal or no signal results in attraction only, an individual then moves towards the center of the group within its maximum interaction zone
\item a weakening signal results in moderate attraction and strong alignment with neighbours. It is this intermediate response which allows net movement along the filament and prevents a low polarity swarm from forming
\item if an improving signal is being experienced this is the optimal direction, interaction zones are reduced to zero, all neighbours are ignored and current direction is maintained.
\end{itemize}
Qualitatively similar results to those presented here were obtained using different functions for each scaling factor so long as the dynamical responses to the signal listed were approximated.
\par

By following the type of rules outlined an individual acting within a group is able to effectively track a filament and locate the source of the signal be it a nutrient source, a potential mate or the location of a suitable habitat. Sequential snapshots of the search behavior being perfomed are shown in Fig.~\ref{fig1} for 60 individuals, while animations are included as supplementary material.

\section{Group size effects}
To further quantify this behavior and understand the role of the parameters in the model we investigate first the effects of the active group size on the probability of successfully locating the source. We refer here to active group size as the number of individuals present at the beginning of each simulation and ignore the potential for populations to fracture as the search is performed. Numerical simulations were performed with a given number of individuals positioned on the chemical filament with a random intial orientation distributed around the half circle pointing towards the source. The number of individuals able to locate the source were recorded and the probabilities of success calculated for varying parameter values. It should be noted that due to the lack of significant persistent gradient parallel to the desired direction navigating a filament upstream or downstream is almost equivalent hence the anisotropy of the initial conditions. However it can be assumed any organism wishing to track a filament will have rudimentary knowledge of the flow direction and this may even be passively obtained through re-orientation by the rheology of the flow.
\par
Fig.~\ref{fig2}a shows how success rate is affected for various group sizes when the desired inter-individual spacing is changed. From this we can see that this distance is potentially more than an avoidance mechanism as it allows the group to act as an efficient network of sensors independently sampling the spatiotemporally evolving environment. To examine this further we rescale the rates of success in terms of the effective area occupied by the group. The effective area is defined as the ideal exclusion area an individual wishes to maintain multiplied by the number of individuals in the population. The result of this rescaling is included in Fig.~\ref{fig2}b. This shows the existence of an optimum area with a sharp increase in success as this is reached then a slow exponential decay as the area is overtaken. These results suggest that although the optimal area is invariant maximum success is somewhat improved with larger numbers, large populations tightly packed together are more robust and better able to track these filaments. 
\par
However this comes at a cost our simulations show the percentage of the group which will arrive at the source on a successful trial will reduce as total group size is increased. For larger groups there is a fission effect and groups will split with some losing the signal and moving out of range of the successful group. To maintain group cohesion the number of individuals present cannot be too large. This may not be a disadvantage in some robotic search applications where the proportion of group reaching the target is unimportant.

\section{The role of memory}
The decay rate, $\alpha$ of the stored concentration value (to which current values are compared) represents the length of an individual's memory. This allows comparison of experienced concentration levels and the assessment of the current trajectory. The value of $\alpha$ defines a timescale for the memory decay and this effectively controls the sensitivity to environmental conditions and represents an individual level parameter which can be tuned through evolution or experience to affect the success of the group. 
\par
In the flow regime we are considering the filamental structure of the advected chemical signal means that the largest gradient is transverse to the desired direction. The memory parameter therefore controls how responsive an individual is to reduction in signal strength as it moves towards the edges of the filament. To place the timescale of the memory in the context of the properties of the signal we approximate the memory parameter required to effectively trace the edges of the filament. By measuring the average width of the filament and assuming its profile follows a Gaussian distribution we calculate the time taken to traverse one standard deviation travelling at $45^{\circ}$ to the parallel direction as \~{}$0.0125$. As the role of memory is to allow an individual to determine if conditions are improving or deteriorating this value defines a memory parameter when it can be assumed an individual is able to respond as it moves towards the edges of the filament. 
\par
Decay rates which are greater than \~{}$0.0125$ lead to less responsiveness and more binary behavior, individuals ignore others when experiencing any concentration value and then attract when the filament is lost. In the limiting case of no memory this corresponds to a low polarization swarm of the type seen in groups of mosquitoes or midges in which there is no net motion but the group is always able to maintain its position centred on the plume. 
\par 
In the other extreme high values of $\alpha$ lead to fluctuations in concentration having a larger impact on success rates and weakening responsiveness to the decay of the filament perpendicular to the desired direction. Effectively individuals that have experienced a high concentration parcel are subsequently less able to accurately respond to the horizontal profile of the filament within the time frame defined by the decay rate. The negative impact of longer memory is a result of intermittent fluctuations in the signal and is therefore a weak effect, meaning once the required memory is reached success rate declines slowly as it is increased.
\par
\begin{figure}[h]
\includegraphics[angle=-90,scale=0.33,clip]{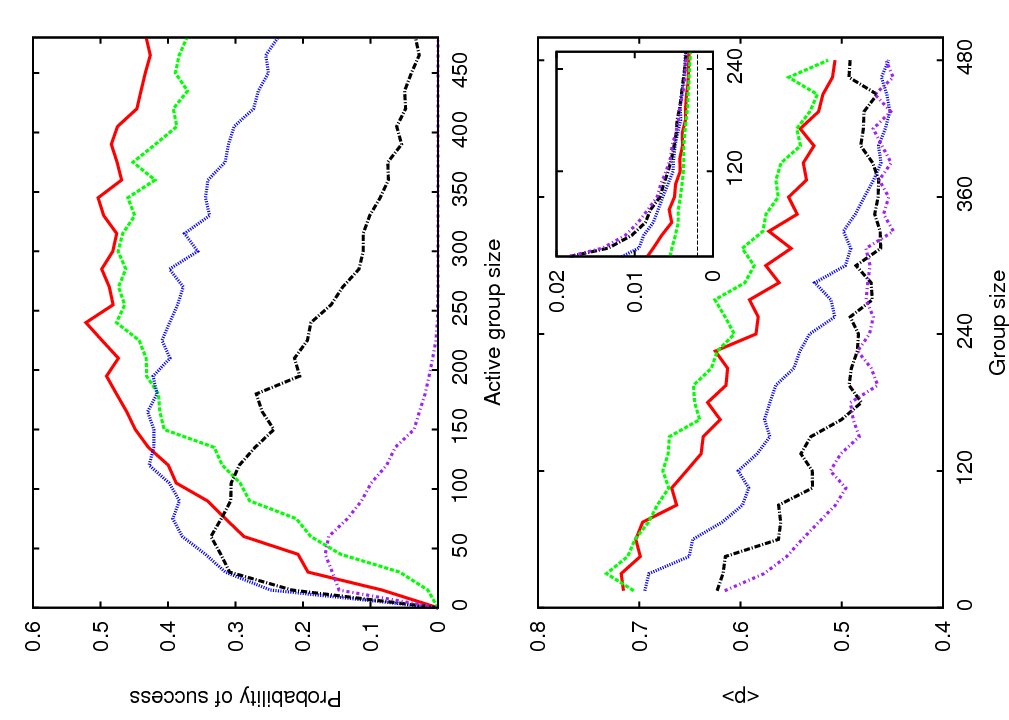}
\caption{(a) Active group size vs success rate for various memory lengths, green dash $\alpha=25\times10^{-3}$, red solid $\alpha=12.5\times10^{-3}$, blue dot $\alpha=2.5\times10^{-3}$, black dot-dash $\alpha=1.25\times10^{-3}$, purple dash-dot $\alpha=0.5\times10^{-3}$. (b) Active group size vs group polarity and (inset) mean nearest neighbout distance. Legend as previous, both these values represent ensemble average values, time averaged within and across multiple simulation instances.}
\label{fig3}
\end{figure}

To examine the intermediate regime the value of $\alpha$ was varied and success rates recorded for different group sizes. These results are shown in Fig.~\ref{fig3}a, from which it is clear optimum memory length is dependent on group size. While the global optimum value corresponds to the value required to detect the edges of the filament while also being able to forget high concentration patches, different group sizes do not all share this same optimum. The reason for this can be found in the tradeoff between exploration and exploitation of the signal where exploration means the effective sampling of the signal while exploitation corresponds to net motion towards its source. We illustrate this by measuring two properties of the group dynamics, the average nearest neighbour distance
\begin{equation}
\langle \Delta r \rangle = \frac{1}{N} \sum_{i=1}^{N} \min_{j}\left(\vert\mathbf{r}_j(t) - \mathbf{r}_i(t) \vert \right)
\end{equation}
and the average group polarity
\begin{equation}
\langle p \rangle = \frac{1}{N} \Big\vert {\sum_{i=1}^{N} \mathbf{p}_i(t)} \Big\vert 
\end{equation}
which are measured numerically and shown in Fig.~\ref{fig3}b. It can be seen the fast decay of memory (which causes more asocial behavior until concentration is greatly reduced) leads to consistently lower polarity and also for smaller groups to be less densely packed. For smaller populations becoming too sensitive to their environment leads to a more highly polarized, compact grouping which is less stable and is more likely to lose track of the filament when required to respond quickly to curves generated by vortices in the flow. This may be compensated for by decreasing $\alpha$ which results in groups covering a larger spatial area, alignment being reduced and results in less tendency to collectively lose the filament. The polarity of the group defines the speed in which it will travel towards the source but the optimum value is a trade-off between speed and exploring the surrounding area which enables the group as a whole to move in different directions when needed.

\section{Discussion}
Our model illustrates how simple adaptive social interactions can lead to cooperative behavior that in this case produces an emergent group level search. By modifying their behavior based on local conditions autonomous individuals enable the group to collectively act as a spatial non-local gradient sensor able to track a chemical signal and locate its source. It has previously been shown how small numbers of individuals can control decision making \cite{nature} and lead groups in a given direction. Here the context-dependent interactions mean leaders are dynamically changing and automatically selected. As these leaders are those which are experiencing increases in local concentration group direction is towards the source.
\par

Although we have chosen to study an advection dominated regime where chaotic dynamics lead to patchy, heterogeneous information the mechanism described here is equally applicable to diffusion dominated environments where smoother chemical values increase steadily as their source is approached. In this case group dynamics can be easily seen to follow a trajectory towards the source however asocial strageties such as a biased random walk \cite{berg} are equally effective in these conditions and social behavior is redundant (even maladaptive due to effects of competition). The interesting case from both an evolutionary and technological perspective is when the nature of the environment renders individual strategies ineffective. 
\par
In the case of the model considered in this paper the center of the filament can be considered as an unstable position for any organism to maintain. That the flow also advects the individuals in our model is both beneficial as it aids in following the signal and disadvantageous as a particle is rapidly moved away if the signal is lost. A strategy based on temporal sampling and biased random turns in these environments is ineffective as once the signal is lost it becomes highly difficult to relocate. Responding to the edges of the filament through a counter-turning mechanism may be effective and is observed in nature \cite{grasso} but requires spatially separate sensors to  accurately determine turning speed required \cite{grasso2}. The advantage of sociality therefore lies in two factors, firstly the stability it provides in maintaining position on the signal. Although lone individuals are easily dislodged and moved away from the signal as a result of exponentially diverging tragectories, collective interactions anchor the group to the filament. Secondly the group is able to sample over a much larger area simultaneously and due to the context-dependent interactions makes a consensus decision based on these samples as if it was a single organism. This results in the smoothing out of fluctuations in the chemical concentration, a comparison of different trajectories without losing track of the signal and the automatic selection of the optimal path. The parameter values and the exact form of the response functions could be tuned further by some evolutionary selection or learning process, to produce optimal behavior adapted to various environments with different statistical properties of the fluctuations of the information carrying signal.
\par
The recent study of animal groups as complex adaptive systems has shown they are able to perform sophisticated tasks through simple local interactions. Aside from the interest in these systems from an evolutionary or biological perspective this work can also have applications in the development of biomimetic technology. Distributed systems have many advantages over the traditional paradigm of centralized control, notably the absence of communication time, inherent robustness (any given component is expendable) and cost-effectiveness as large numbers of simple robotic components can be manufactured efficiently. In the traditional sense the results outlined in this paper may be applied to robotic search strategies but may also be applied to more abstract search problems in the domain of genetic algorithms e.g. particle swarm optimization as our model illustrates mechanisms of this type can be effective even in the absence of global information. 

\par

\par




\section{materials}
The isotropic flow field was generated by the stochastic evolution in Fourier space of an exponentially decaying energy spectrum with the lengthscale of the highest energy modes set to 0.31. Root mean squared velocity was 0.25 and a mean flow of 0.6 was imposed in a constant direction. Advection of the chemical signal was performed using a semi-Lagrangian method with a gridsize of 512. 
Particle positions were updated using a second order scheme with positions and velocities of neighbors updated once per timestep. Particle speed $v_s$= 1.6, turning rate $\gamma$=140 (radians), dt=0.00025, maximum orientation radius $R_{O,max}$ = 0.075, maximum attraction radius $R_{A,max}$ = 0.125. Success rates were obtained using 1000 trials and recording number of individuals reaching within a distance of 0.025 of the source location. Individuals were initially located in ball centered on the filament at a distance of 0.8 from the source.

\end{document}